\documentclass[12pt]{article}

\usepackage[centertags]{amsmath}
\usepackage{amsfonts}
\usepackage{amssymb}
\usepackage{amsthm}
\usepackage{newlfont}
\usepackage{epsfig}
\usepackage{amscd}

\newcommand{\RR}{{\mathbb R}}

\newcommand{\beq}{\begin{equation}}
\newcommand{\eeq}{\end{equation}}
\newcommand{\ba}{\begin{array}}
\newcommand{\ea}{\end{array}}
\newcommand{\bea}{\begin{eqnarray}}
\newcommand{\eea}{\end{eqnarray}}

\begin{document}

\begin{center}
{\Large \sc \bf Inverse Scattering Problem} 

\vskip 5pt
{\Large \sc \bf for Vector Fields}

\vskip 5pt
{\Large \sc \bf and the Heavenly Equation} 

\vskip 30pt

{\large  S. V. Manakov$^{1,\S}$ and P. M. Santini$^{2,3,\S}$}

\vskip 30pt

{\it 
$^1$ Landau Institute for Theoretical Physics, Moscow, Russia

\smallskip

$^2$ Dipartimento di Fisica, Universit\`a di Roma "La Sapienza"\\
Piazz.le Aldo Moro 2, I-00185 Roma, Italy

\smallskip

$^3$ Istituto Nazionale di Fisica Nucleare, Sezione di Roma 1\\
P.le Aldo Moro 2, I-00185 Roma, Italy}

\bigskip

$^{\S}$e-mail:  {\tt manakov@itp.ac.ru, paolo.santini@roma1.infn.it}

\bigskip

{\today}

\end{center}

\begin{abstract}
We solve the inverse scattering problem for multidimensional vector fields 
and we use this result to construct the formal solution of the Cauchy problem for  
the second heavenly equation of Plebanski, a scalar nonlinear partial differential equation in four  
dimensions relevant in General Relativity, which arises from the commutation of multidimensional 
Hamiltonian vector fields. 
\end{abstract}
\section{Introduction}
In this paper we solve the inverse scattering problem for 
multidimensional vector fields and we use this result to construct  
the formal solution of the Cauchy problem for the real second heavenly equation
\beq
\label{heavenly1}
\theta_{,tx}-\theta_{,zy}+\theta_{,xx}\theta_{,yy}-\theta^2_{,xy}=0,~~~
\theta=\theta(x,y,z,t)\in\RR,~~~~~x,y,z,t\in\RR,
\eeq 
where $\theta_{,x}=\partial\theta/\partial x,~\theta_{,xy}=\partial^2\theta/\partial x\partial y$. 

This scalar nonlinear partial differential equation (PDE) in $4$ independent variables 
$x,y,z,t$, introduced in \cite{Pleb} by Plebanski, describes the Einstein field equations that 
govern self-dual gravitational fields. Its 
$2$-dimensional reduction $\theta_{,z}=\theta_{,t}=0$ is the Monge-Amp\`ere 
equation, relevant in Differential Geometry. As we shall see in the following, 
the heavenly equation plays also a distinguished role in the theory of commuting 2-dimensional, Hamiltonian 
dynamical systems. 
 
The heavenly equation, together with the 
equations for the self-dual (and anti-self-dual) Yang-Mills (SDYM) fields \cite{YM}, are perhaps the most distinguished 
examples of nonlinear PDEs in more than three independent variables arising as commutativity 
conditions of linear operators \cite{ZS}, and therefore amenable, in principle, 
to exact treatments based on the spectral theory of those operators \cite{ZMNP},\cite{AC}. If the SDYM equations 
are considered on an abstract Lie algebra, then the heavenly equation can actually be interpreted as a distinguished 
realization of the SDYM equations, corresponding to the Lie algebra of divergence free vector fields independent of 
the SDYM coordinates \cite{CMN}.  Equation (\ref{heavenly1}) has been investigated within the twistor approach  
in \cite{DuMa}. A bi-Hamiltonian formulation, a hodograph transformation, and a nonlinear dressing formalism for equation 
(\ref{heavenly1}) have been recently constructed, respectively, in \cite{NNS}, \cite{MM} and \cite{BK}. 

We end this introductory section remarking that the inverse scattering problem for vector fields 
developed in this paper presents some 
interesting new features with respect to the traditional Inverse Scattering Transform (IST) in multidimensions. For instance: 

\begin{itemize}
\item The space of eigenfunctions of the multidimensional vector fields is a ring. In the heavenly reduction, 
the space of eigenfunctions is not only a 
ring, but also a Lie algebra, with Lie bracket given by the  natural Poisson bracket.

\item Constructing solutions of the integrable multidimensional PDEs arising from the commutation 
of vector fields is equivalent to characterizing commuting pairs of dynamical systems, Hamiltonian 
in the heavenly reduction.

\item The scattering data are intimately related to the scattering vector $\vec\Delta$, associated with the dynamical 
system which describes the characteristic curves of the vector field. These scattering data, in one to one 
correspondence with the coefficients of the vector field, depend on 
a number of spectral variables equal to the number of space variables. Thus the counting is consistent (a highly 
non generic feature in multidimensional IST).

\item A central role in this IST is played by a certain Riemann-Hilbert problem with a shift depending on the scattering 
vector $\vec\Delta$.

\item The above ring property allows one to construct two essentially different IST schemes. The first one, in terms 
of a set of eigenfunctions intimately related to the integral curves of the associated vector field, 
is characterized by a nonlinear inverse problem. The second one, involving more traditional Jost and 
analytic eigenfunctions, obtained ``exponentiating'' the above eigenfunctions, is instead characterized by 
a linear inverse problem. 
   
\end{itemize} 

\section{The Cauchy problem}

\noindent
{\bf A multidimensional Lax pair}. It is known that the commutation of multidimensional vector fields  
leads to nonlinear first order multidimensional PDEs (see, f.i., \cite{ZS}). 

Consider, for instance, the pair of operators
\beq
\label{L1L2}
\hat L_i=\partial_{t^i}+\lambda\partial_{z^i}+
\sum\limits_{k=1}^Nu^k_i\partial_{x^k}=\partial_{t^i}+\lambda\partial_{z^i}+\vec u_i\cdot\nabla_{\vec x},~~~~i=1,2 
\eeq  
where $\partial_x$ denotes partial differentiation with respect to the generic variable $x$, $\vec x=(x^1,..,x^N)$,  
$\nabla_{\vec x}=(\partial_{x^1},..,\partial_{x^N})$, $\vec u_i=(u^1_i,..,u^N_i),~i=1,2$, 
$\lambda$ is a complex parameter and the vector coefficients $\vec u_i$ depend on the 
independent variables $t^i,z^i,x^k$, $i=1,2,~k=1,..,N$, but not  
on $\lambda$. The existence of a common eigenfunction $f$ for the operators $\hat L_1$ and $\hat L_2$:
\beq
\label{Lax1}
\hat L_1f=\hat L_2f=0,
\eeq
implies their commutation, $\forall\lambda$:
\beq
[\hat L_1,\hat L_2]=0,
\eeq  
which is equivalent to the following system of $2N$ first order quasi-linear PDEs in $(4+N)$ dimensions:
\beq
\label{quasilin-u}
\ba{l}
u^{k}_{1,z^2}=u^{k}_{2,z^1},~~~~~~~~~~k=1,..,N, \\
u^{k}_{1,t^2}-u^{k}_{2,t^1}+
\sum\limits_{l=1}^N\left(u^{l}_2 u^{k}_{1,x^{l}}-u^{l}_1u^{k}_{2,x^{l}}\right)=0,~~~~~~~~~~k=1,..,N.
\ea
\eeq
Parametrizing the first set of equations in terms of the potentials $U^{k}$ 
\beq
u^{k}_i=U^{k}_{,z^i},~~~~~~i=1,2,~~~k=1,..,N,
\eeq
one obtains the following system of $N$ nonlinear 
PDEs for the $N$ dependent variables $U^{k}$ in $(4+N)$ dimensions:
\beq
\label{quasilin-U}
U^{k}_{,t^1z^2}-U^{k}_{,t^2z^1}+
\sum\limits_{l=1}^N\left(U^{l}_{,z^1}U^{k}_{,z^2x^l}-U^{l}_{,z^2}U^{k}_{,z^1x^l}\right)=0,
~~~~k=1,..,N.
\eeq
This system admits a natural reduction; indeed, applying the operator $\sum_{k=1}^N\partial_{x^{k}}$ to equations 
(\ref{quasilin-U}) one obtains
\beq
\left[{\partial}_{t^1}{\partial}_{z^2}-{\partial}_{t^2}{\partial}_{z^1}+
\sum\limits_{k=1}^N\left({U^{k}}_{,z^1}{\partial}_{z^2}{\partial}_{x^k}-
{U^{k}}_{,z^2}{\partial}_{z^1}{\partial}_{x^k}\right)\right]
\sum\limits_{l=1}^NU^{l}_{,x^{l}}=0,
\eeq
from which one infers that the condition
\beq
\label{constraint-U}
\sum\limits_{k=1}^NU^{k}_{,x^{k}}=0
\eeq
is an admissible reduction for equation (\ref{quasilin-U}), implying that the condition of zero-divergence: 
\beq
\label{red-u}
\nabla_{\vec x}\cdot\vec u_i=0,~~i=1,2
\eeq
is an admissible constraint for the vectors $\vec u_i,~i=1,2$. 

From now on, we concentrate our attention on the following important example:
\beq
\label{N=2}
N=2,~~~z^i=x^i,~~~i=1,2.
\eeq
In this case, the zero-divergence reduction (\ref{red-u}) makes the two vector fields \break 
$\vec u_i\cdot\nabla_{\vec x}$ 
Hamiltonian, allowing for the introduction of two Hamiltonians $H_i,~i=1,2$ such that:
\beq
u^{j}_i=\epsilon^{jk}{H_i}_{,x^k},\;\;\;i,j,k=1,2,
\eeq
where $\epsilon^{jk}$ is the totally anti-symmetric tensor, which, due to (\ref{quasilin-u}a), 
are parametrized by a single potential $\theta$:
\beq
\ba{l}
H_i=\theta_{,x^i}\;\;
\left(u^j_i=\epsilon^{jk}{H_i}_{,x^k}=\epsilon^{jk}\theta_{,x^ix^k},~~
U^i=\epsilon^{jk}H_j=\epsilon^{ij}{\theta}_{.x^j},\;\;\;i,j=1,2\right).
\ea
\eeq
Then the compatible linear problems (\ref{Lax1}) can be written down as Hamilton equations with respect to the times $t^1,t^2$:
\beq
\label{HamEqu}
\ba{l}
f_{,t^1}=\{H_1+\lambda x^2,f\}_{\vec x}, \\
f_{,t^2}=\{H_2-\lambda x^1,f\}_{\vec x}, 
\ea
\eeq
where $\{\cdot,\cdot\}_{\vec x}$ is the Poisson bracket with respect to the variables $x^1,x^2$:
\beq
\label{PB}
\{f,g\}_{\vec x}=f_{,x^1}g_{,x^2}-f_{,x^2}g_{,x^1},
\eeq
and the nonlinear system (\ref{quasilin-U}) reduces to the heavenly equation 
in Hamiltonian form  
\beq
\label{heavenly2}
\theta_{,t^2x^1}-\theta_{,t^1x^2}+\{\theta_{,x^1},\theta_{,x^2}\}_{\vec x}=\mbox{constant},~~~
\eeq 
equivalent to (\ref{heavenly1}) after choosing the constant to be zero, with the following 
identification of the variables
\beq
\label{xyzt}
t^1=z,~t^2=t,~~x^1=x,~~x^2=y,
\eeq
which we are going to make in the rest of the paper.

\vskip 10pt
\noindent
{\bf Commuting dynamical systems}. It is well-known (see, f.i., \cite{CH}) that linear first order PDEs 
like (\ref{Lax1}) are intimately related to systems of ordinary differential equations describing their 
characteristic curves. The dynamical 
systems associated with the vector fields $\hat L_{1,2}$:
\beq
\label{L1L2bis}
\ba{l}
\hat L_1=\partial_{z}+\lambda\partial_{x}+\vec u_1\cdot\nabla_{\vec x}, \\
\hat L_2=\partial_{t}+\lambda\partial_{y}+\vec u_2\cdot\nabla_{\vec x},
\ea
\eeq
where $\vec x=(x,y)$  and $\nabla_{\vec x}=(\partial_x,\partial_y)$, are: 
\beq
\label{dynsyst}
\ba{l}
\hat L_1 :\;\;\;\frac{d\vec x}{dz}=\vec u_1(\vec x,z)+(\lambda,0), \\
~~  \\
\hat L_2 :\;\;\;\frac{d\vec x}{dt}=\vec u_2(\vec x,t)+(0,\lambda).
\ea
\eeq

We remark that the two flows generated by the times $z$ and $t$ commute, $\forall\lambda$,  iff the 
fields $u^j_i$  satisfy the integrable quasi-linear equations (\ref{quasilin-u}),(\ref{N=2}). Therefore:

\noindent
{\it Constructing solutions of the integrable PDEs (\ref{quasilin-u}),(\ref{N=2})  is equivalent to 
solving the classical problem of constructing pairs of commuting 2-dimensional dynamical 
systems}.

We also remark that, in the heavenly zero-divergence reduction (\ref{red-u}), 
the commuting flows (\ref{dynsyst}) are Hamiltonian:
\beq
\label{dynsyst2}
\ba{l}
\frac{d\vec x}{dz}=\{\vec x,H_1+\lambda y\}_{\vec x}, \\
~~  \\
\frac{d\vec x}{dt}=\{\vec x,H_2-\lambda x\}_{\vec x}.
\ea
\eeq 

Consider now a solution $(r^1,r^2)$ of equation (\ref{dynsyst}a) and assume that $u^j_1\to 0$ as $|z|\to\infty$;  
then the phase space point $\vec r=(r^1,r^2)$ travels asymptotically with constant speed $(\lambda,0)$: 
\beq
\label{r-asympt}
\vec r\to\vec s_{\pm}+(\lambda,0)z,~~~~~z\to\pm\infty,
\eeq    
where $\vec s_{\pm}$ are constant vectors. 

If $\vec s_-$ is given, then $\vec s_+$ can be viewed as a function of $\vec s_-$ and $\lambda$: $\vec s_+(\vec s_-,\lambda)$,  
and the difference $\vec\Delta$ between these asymptotic positions 
\beq
\label{Delta}
\vec\Delta(\vec s_-,\lambda)=\vec s_+-\vec s_-
\eeq
describes the $z$-scattering of the phase space point of the dynamical system (\ref{dynsyst}a).  
As we shall see in the following, the scattering vector  
$\vec\Delta$ is also intimately related to the $z$-scattering datum $S$ associated with the partial 
differential operator $\hat L_1$ (see (\ref{FT-S})), which plays a central role in the IST  
for the vector field $\hat L_1$, and, consequently,  in the solution of the Cauchy problem 
for the nonlinear PDEs (\ref{quasilin-u}),(\ref{N=2}), and for the heavenly reduction (\ref{heavenly1}).

Now we consider such a Cauchy problem,    
within the class of rapidly decreasing real potentials $u^j_i$:
\beq
\label{localization}
\ba{l}
u^j_i\to~0,~~(x^2+y^2+z^2)\to\infty, \\
u^j_i\in\RR,~~~(x,y,z)\in\RR^3,~~t>0,
\ea
\eeq
interpreting $t$ as time and the other three variables $x,y,z$ as space variables.  

To solve such a Cauchy problem by the IST method (see, f.i., \cite{ZMNP},\cite{AC}), we construct the 
IST for the operator $\hat L_1$ in (\ref{L1L2bis}a), within the class of rapidly decreasing real potentials, 
interpreting the operator $\hat L_2$ in (\ref{L1L2bis}b) as the time operator. 

\vskip 5pt
\noindent
{\bf Eigenfunctions and their properties}. Since $\hat L_{1,2}$ are linear, first order, 
partial differential operators with scalar coefficients, the product of two solutions $f_{1,2}$ of the 
equation $\hat L_if=0$ is also a solution: $\hat L_i(f_1f_2)=0$ (in general, an arbitrary function $G(f_1,f_2)$ 
of these two solutions is also a solution: $\hat L_iG(f_1,f_2)=0$). Therefore the space of solutions of the equation 
$\hat L_if=0$ forms a ring. As we shall see in the following, this property introduces important novelties in the 
Inverse Scattering formalism.   

The localization (\ref{localization}) of the 
vector potential $\vec u_1$ implies that, if $f$ is a solution of $\hat L_1 f=0$, then 
\beq
\label{asymptf}
\ba{l}
f(\vec x,z,\lambda)\to f_{\pm}(\vec X,\lambda),\;\;z\to\pm\infty, \\
\vec X:=\vec x-(\lambda,0)z;
\ea
\eeq
i.e., asymptotically, $f$ is an arbitrary function of $(x-\lambda z)$, $y$ and $\lambda$.

A central role in the theory is played by the real vector eigenfunctions $\vec\varphi_{\pm}(\vec x,z,\lambda)$, the solutions of 
$\hat L_1\vec\varphi_{\pm}=\vec 0$ defined by the asymptotics 
\beq
\label{asymptvarphi}
\vec\varphi_{\pm}(\vec x,z,\lambda)\to \vec X,\;\;\;z\to\pm\infty.
\eeq   
Their connection to the dynamical system (\ref{dynsyst}a) is immediate. Let $\vec x=\vec r_{\pm}(\vec\omega,z,\lambda)$ 
be the solutions of the dynamical system (\ref{dynsyst}a) satisfying:
\beq
\label{dyn}
\ba{l}
\frac{d\vec r_{\pm}}{dz}=\vec u_1(\vec r_{\pm},z)+(\lambda,0), \\
\vec r_{\pm}(\vec\omega,z,\lambda)\to \vec\omega +(\lambda,0)z,~~z\to\pm\infty,
\ea 
\eeq
where $\vec\omega\in\RR^2$ is a given constant. Then the vector eigenfunctions $\vec\varphi_{\pm}$ arise from 
solving the equation $\vec x=\vec r_{\pm}(\vec\omega,z,\lambda)$ with respect to $\vec\omega$:
\beq
\label{rel_rvarphi}
\vec x=\vec r_{\pm}(\vec\omega,z,\lambda)~~\Leftrightarrow~~\vec\omega=\vec\varphi_{\pm}(\vec x,z,\lambda).
\eeq
Hereafter, when systems of equations are inverted, we will always consider a neighborhood of a point in which 
the corresponding Jacobian does not vanish.

Due to the ring property of the space of eigenfunctions, an arbitrary function of $\vec\varphi_{\pm}$ is also 
an eigenfunction. The Jost eigenfunctions $\phi^{\pm}$ are defined by:
\beq
\label{phi-exp}
\phi^{\pm}(\vec x,z;\vec\alpha,\lambda):=e^{i\vec\alpha\cdot\vec\varphi_{\pm}(\vec x,z,\lambda)}.
\eeq   
They satisfy the asymptotics:
\beq
\label{Jost}
\phi^{\pm}(\vec x,z;\vec \alpha,\lambda)\to e^{i\vec\alpha\cdot\vec X},~~~z\to\pm\infty,
\eeq
where $\vec\alpha=(\alpha_1,\alpha_2)$ are arbitrary real parameters and $\vec\alpha\cdot\vec X$ is the usual scalar product 
in $\RR^2$. They are equivalently characterized by the integral equations
\beq
\label{phi}
\ba{l}
\phi^{\pm}(\vec x,z;\vec \alpha,\lambda)+ \\
\int_{\RR^3}d{\vec x}'dz'G^{\pm}(\vec x-\vec x',z-z';\lambda)\left(\vec u_1(\vec x',z')\cdot\nabla_{\vec x'}\right)
\phi^{\pm}(\vec x',z';\vec \alpha,\lambda)=e^{i\vec\alpha\cdot\vec X},
\ea
\eeq
in terms of the Green's functions
\beq
G^{\pm}(\vec x,z;\lambda)=\mp\theta(\mp z)\delta(x-\lambda z)\delta(y).
\eeq

A crucial role in inverse scattering is also played by analytic eigenfunctions. In the IST for 
the vector field $\hat L_1$,  
the analytic eigenfunctions $\psi_{\pm}(\vec x,z;\vec \alpha,\lambda)$ are the solutions of $\hat L_1\psi_{\pm}=0$ 
satisfying the integral equations 
\beq
\label{psi}
\ba{l}
\psi_{\pm}(\vec x,z;\vec \alpha,\lambda)+ \\
\int_{\RR^3}d{\vec x}'dz'G_{\pm}(\vec x-\vec x',z-z';\lambda)
e^{i\vec\alpha\cdot(\vec X-\vec X')}\left(\vec u_1(\vec x',z')\cdot\nabla_{\vec x'}\right)
\psi_{\pm}(\vec x',z';\vec \alpha,\lambda)= 
e^{i\vec\alpha\cdot\vec X},
\ea
\eeq
where $G_{\pm}$ are the analytic Green's functions
\beq
\label{Green_analytic}
G_{\pm}(\vec x,z;\lambda)=\pm\frac{\delta(y)}{2\pi i[x-(\lambda\pm i\epsilon) z]}.
\eeq
The analyticity properties of $G_{\pm}(\vec x,z;\lambda)$ in the complex $\lambda$-plane 
imply that $\psi_{+}(\vec x,z;\vec \alpha,\lambda)e^{-i\vec\alpha\cdot\vec X}$ and 
$\psi_{-}(\vec x,z;\vec \alpha,\lambda)e^{-i\vec\alpha\cdot\vec X}$ are 
analytic, respectively, in the upper and lower half $\lambda$ - plane, with 
the following asymptotics, for large $\lambda$:
\beq
\label{asympt1}
\psi_{\pm}(\vec x,z;\vec \alpha,\lambda)e^{-i\vec\alpha\cdot\vec X}=
1+\frac{\vec\alpha\cdot \vec Q_{\pm}(\vec x,z)}{\lambda}+O(\lambda^{-2}),~~~|\lambda|>>1,
\eeq
where:
\beq
\label{asympt2}
\vec Q_{\pm}(\vec x,z)=\pm P\int_{\RR^2}\frac{dx'dz'}{2\pi(z-z')}\vec u_1(x',y,z')-
\frac{i}{2}\left(\int\limits_{-\infty}^x-\int\limits_{x}^{\infty}\right)dx'\vec u_1(x',y,z),
\eeq
entailing that
\beq
\label{u-Q}
\vec u_1(\vec x,z)=i\vec {Q_{\pm}}_{,x}(\vec x,z).
\eeq

It is important to remark that the analytic Green's functions (\ref{Green_analytic}) exhibit the following 
asymptotics for $z\to\pm\infty$:
\beq
\ba{l}
G_{\pm}(\vec x,z;\lambda)\to\pm\frac{\delta(y)}{2\pi i[x-\lambda z \mp i\epsilon]},\;\;z\to +\infty, \\
G_{\pm}(\vec x,z;\lambda)\to\pm\frac{\delta(y)}{2\pi i[x-\lambda z \pm i\epsilon]},\;\;z\to -\infty,
\ea
\eeq
entailing that the $z=+\infty$ asymptotics of $\psi_{+}$ and $\psi_{-}$ are analytic 
respectively in the lower and upper halves of the complex plane $(x-\lambda z)$, while the $z=-\infty$ 
asymptotics of $\psi_{+}$ and 
$\psi_{-}$ are analytic respectively in the upper and lower halves of the complex plane $(x-\lambda z)$ (similar features 
were obtained in \cite{MZ}).  

Since the solution space of the equation $\hat L_1f=0$ is a ring, then the product of two 
Jost (analytic) solutions corresponding to different vector parameters $\vec\alpha,\vec \beta\in\RR^2$  
is still a Jost (analytic) solution satisfying:
\beq
\label{functphi}
\ba{l}
\phi^{\pm}(\vec x,z;\vec \alpha,\lambda)\phi^{\pm}(\vec x,z;\vec \beta,\lambda)=
\phi^{\pm}(\vec x,z;\vec \alpha+\vec \beta,\lambda), \\
\psi_{\pm}(\vec x,z;\vec \alpha,\lambda)\psi_{\pm}(\vec x,z;\vec \beta,\lambda)=
\psi_{\pm}(\vec x,z;\vec \alpha+\vec \beta,\lambda).
\ea
\eeq
Choosing $\vec\alpha=(1,0)$ and $\vec\beta=(0,1)$, equations (\ref{functphi}) imply that the Jost (analytic) eigenfunctions 
depend on $\vec\alpha$ only at the exponents:
\beq
\label{phi-alpha-dep}
\ba{l}
\phi^{\pm}(\vec x,z;\vec \alpha,\lambda)=
\left(\phi^{\pm}_1(\vec x,z;\lambda)\right)^{\alpha_1}\left(\phi^{\pm}_2(\vec x,z;\lambda)\right)^{\alpha_2}=
e^{i\vec\alpha\cdot\vec\varphi_{\pm}(\vec x,z;\lambda)}, \\
\psi_{\pm}(\vec x,z;\vec \alpha,\lambda)=
\left({\psi_{\pm}}_1(\vec x,z;\lambda)\right)^{\alpha_1}\left({\psi_{\pm}}_2(\vec x,z;\lambda)\right)^{\alpha_2}=:
e^{\vec\alpha\cdot\vec\pi_{\pm}(\vec x,z;\lambda)},
\ea
\eeq
where ${\psi_{\pm}}_1(\vec x,z;\lambda)$ and ${\psi_{\pm}}_2(\vec x,z;\lambda)$ are the analytic eigenfunctions 
satisfying the integral equations (\ref{psi}) for $\vec\alpha=(1,0)$ and $\vec\alpha=(0,1)$ respectively. 

Due to the ring property, the vector functions $\vec\pi_{\pm}$, appearing in (\ref{phi-alpha-dep}b) and defined by
\beq
\vec\pi_{\pm}(\vec x,z,\lambda):=\big(\log{\psi_{\pm}}_1(\vec x,z;\lambda),\log{\psi_{\pm}}_2(\vec x,z;\lambda)\big),
\eeq
are also eigenfunctions of $\hat L_1:~\hat L_1\vec\pi_{\pm}=\vec 0$. In addition, since $\psi_{\pm}$ are 
analytic in $\lambda$, $\forall~\vec\alpha\in\RR^2$, it follows that $\psi_{1,2}$ in (\ref{phi-alpha-dep}b) 
cannot have zeroes and poles in their analyticity domains. Therefore $\vec\pi_+$ and $\vec\pi_-$ are also 
analytic, respectively, in the upper and lower halves of the complex $\lambda$-plane, with asymptotics:
\beq
\label{asympt_pi}
\vec\pi_{\pm}(\vec x,z,\lambda)=i\vec X+\frac{\vec Q_{\pm}(\vec x,z)}{\lambda}+O(\lambda^{-2}).
~~~|\lambda|>>1,
\eeq  
Analogously, the $z=+\infty$ asymptotics of $\vec\pi_{+}$ and $\vec\pi_{-}$ are analytic 
respectively in the lower and upper halves of the complex plane $(x-\lambda z)$, while the $z=-\infty$ asymptotics 
of $\vec\pi_{+}$ and 
$\vec\pi_{-}$ are analytic respectively in the upper and lower halves of the complex plane $(x-\lambda z)$.

Since the vector fields are real ($\vec u_i=\vec u_i^*$), we have the following obvious restrictions for the solutions 
$\phi^\pm$ and $\psi_\pm$ (for real $\lambda$):
\begin{eqnarray}
\label {reality_phipsi}
{\phi^\pm}^*(\vec x,z;\vec\alpha,\lambda) = \phi^\pm(\vec x,z;-\vec\alpha,\lambda),\\
{\psi_\pm}^*(\vec x,z;\vec\alpha,\lambda) = \psi_\mp(\vec x,z;-\vec\alpha,\lambda).
\end{eqnarray}

As we shall see in the following sections, the eigenfunctions $\vec\varphi_{\pm}$ and $\vec\pi_{\pm}$ are the 
basic ingredients of 
an IST involving a nonlinear inverse problem, while the eigenfunctions $\phi^{\pm}$ and $\psi_{\pm}$ are the basic 
ingredients of an IST involving a linear inverse problem.  

\vskip 5pt
\noindent
{\bf Scattering data}. We begin this section observing that the vectors $\vec r_{\pm}$ defined in (\ref{dyn}),  
can be identified with the vector $\vec r$ in (\ref{r-asympt}) setting $\vec s_{\pm}=\vec\omega$; then the asymptotics 
of $\vec r_{\pm}$ at the opposite ends of the $z$ axis can be described in terms of the scattering vector 
$\vec\Delta$ in (\ref{Delta}) as follows:
\beq
\label{asymptr+-}
\ba{l}
\vec r_{-}(\vec\omega,z,\lambda)\to \vec s_+(\vec\omega,\lambda)+(\lambda,0)z,\;\;\;z\to +\infty, \\
\vec r_{+}(\vec\omega,z,\lambda)\to \vec\Omega(\vec\omega,\lambda)+(\lambda,0)z,\;\;\;z\to -\infty,
\ea
\eeq
where
\beq
\label{s+}
\vec s_+(\vec\omega,\lambda)=\vec\omega+\vec\Delta(\vec\omega,\lambda),
\eeq
and $\vec\omega=\vec\Omega(\vec s_+,\lambda)$ is the inverse of the transformation (\ref{s+}), from $\vec\omega$ to  
$\vec s_+(\omega,\lambda)$:
\beq
\label{Omega}
\vec s_+=\vec\omega+\vec\Delta(\vec\omega,\lambda)~~\Leftrightarrow~~\vec\omega=\vec\Omega(\vec s_+,\lambda).
\eeq
Due to the intimate connection (\ref{rel_rvarphi}) between $\vec r_{\pm}$ and $\vec\varphi_{\pm}$, also the asymptotics of 
$\vec\varphi_{\pm}$ at the opposite ends of the $z$ axis are described in terms of $\vec\Delta$:
\beq
\label{asymptvarphi-}
\ba{l}
\vec\varphi_-(\vec x,z,\lambda)\to \vec\Omega(\vec X,\lambda),\;\;z\to +\infty, \\
\vec\varphi_+(\vec x,z,\lambda)\to \vec s_+(\vec X,\lambda),\;\;z\to -\infty,
\ea
\eeq
and, consequently, the expression of $\vec\varphi_+$ in terms of $\vec\varphi_-$ is given by:
\beq
\label{rel_phi+phi-}
\vec\varphi_+(\vec x,z,\lambda)=\vec s_+(\vec\varphi_-(\vec x,z,\lambda),\lambda)=\vec\varphi_-(\vec x,z,\lambda)+
\vec\Delta(\vec\varphi_-(\vec x,z,\lambda),\lambda).
\eeq 
To derive equations (\ref{asymptvarphi-}), it is sufficient to take the $z=\pm\infty$ limit of (\ref{rel_rvarphi}), 
using (\ref{asymptr+-}); equation (\ref{rel_phi+phi-}) follows from the ring 
property and from (\ref{asymptvarphi-}).

The Jost solutions form a natural basis for expanding any eigenfunction of $\hat L_1$.  
The expansion of $\phi^{+}$ in terms of $\phi^{-}$: 
\beq
\label{S}
\phi^{+}(\vec x,z;\vec \alpha,\lambda)=\int_{\RR^2}d\vec\beta S(\vec\alpha,\vec\beta,\lambda)
\phi^{-}(\vec x,z;\vec \beta,\lambda)
\eeq
defines the $z$-scattering datum $S$, which admits the standard integral representation:
\beq
\label{S-int-repr}
S(\vec\alpha,\vec\beta,\lambda)=\delta(\vec\alpha-\vec\beta)+
\int_{\RR^3}d{\vec x}dze^{-i\vec\beta\cdot\vec X}
\left(\vec u_1(\vec x,z)\cdot\nabla_{\vec x}\right)\phi^{+}(\vec x,z;\vec \alpha,\lambda)
\eeq
following from (\ref{phi}), where $\delta(\vec\alpha-\vec\beta)=\delta(\alpha_1-\beta_1)\delta(\alpha_2-\beta_2)$. 

The $z$-scattering datum $S$ exhibits the following Fourier representation in terms of the scattering vector 
$\vec\Delta$:
\beq
\label{FT-S}
S(\vec\alpha,\vec\beta,\lambda)=\int_{\RR^2}\frac{d\vec\omega}{(2\pi)^2}e^{i\vec\omega\cdot(\vec\alpha-\vec\beta)+
i\vec\alpha\cdot\vec\Delta(\vec\omega,\lambda)}.
\eeq 
This basic formula is consequence of (\ref{phi-exp}), (\ref{rel_phi+phi-}) and (\ref{S}). Indeed, 
substituting (\ref{phi-exp})$_{\pm}$ into (\ref{S}), one obtains:
\beq
e^{i\vec\alpha\cdot\varphi_+}=\tilde S(\vec\alpha,\varphi_-,\lambda),
\eeq 
where $\tilde S$ is the Fourier transform of $S$, with respect to $\vec\beta$:
\beq
\tilde S(\vec\alpha,\vec\omega,\lambda)=\int_{\RR^2}d\vec\beta 
S(\vec\alpha,\vec\beta,\lambda)e^{i\vec\beta\cdot\vec\omega}.
\eeq 
Then, using (\ref{rel_phi+phi-}), it follows that 
$\tilde S(\vec\alpha,\vec\omega,\lambda)=e^{i\vec\alpha\cdot\vec s_+(\vec\omega,\lambda)}$. 
Therefore the special dependence (\ref{phi-exp}) of $\phi^{\pm}$ on $\vec\alpha$ only at the exponent 
implies that also the Fourier transform 
$\tilde S$ of $S$ depends on $\vec\alpha$ only at the exponent.

The direct problem is the mapping, via (\ref{S-int-repr}), from the two real potentials $u^{1,2}_1$, 
functions of the three real variables $(\vec x,z)$, to the scattering datum $S$, or, via (\ref{FT-S}),    
to the two real components $\Delta^{1,2}$ of $\vec\Delta$, functions of the three real variables $(\vec\omega,\lambda)$. 
Then, thanks to the ring property of the space of eigenfunctions, the counting is consistent.

The expansions of the analytic eigenfunctions $\psi_{\pm}$ in terms of the Jost eigenfunctions:
\beq
\label{K}
\ba{l}
\psi_{+}(\vec x,z;\vec \alpha,\lambda)=
\int_{\RR^2}d\vec\beta K^{\pm}_{+}(\vec\alpha,\vec\beta,\lambda)\phi^{\pm}(\vec x,z;\vec \beta,\lambda), \\
\psi_{-}(\vec x,z;\vec \alpha,\lambda)=
\int_{\RR^2}d\vec\beta K^{\pm}_{-}(\vec\alpha,\vec\beta,\lambda)\phi^{\pm}(\vec x,z;\vec \beta,\lambda)
 \ea
\eeq
introduce the kernels $K^{\pm}_{\pm}(\vec\alpha,\vec\beta,\lambda)$; they are connected to $S$ 
via the integral equations
\beq
\label{K-S}
K^{-}_{\pm}(\vec\alpha,\vec\beta,\lambda)=
\int_{\RR^2}d\vec\gamma K^{+}_{\pm}(\vec\alpha,\vec\gamma,\lambda)S(\vec\gamma,\vec\beta,\lambda),
\eeq
which follow directly from (\ref{K}), replacing $\phi^{+}$ by its expansion (\ref{S}) in terms of $\phi^{-}$. 

The reality of the vector fields and its consequence (\ref{reality_phipsi}) imply the following reality constraints 
for the scattering data $S$ and $K$ (for real $\lambda$):
\begin{eqnarray}
S^*(\vec\alpha,\vec\beta,\lambda)= S(-\vec\alpha,-\vec\beta,\lambda),\\
{K^{\pm}_+}^*(\vec\alpha,\vec\beta,\lambda)= K^{\pm}_-(-\vec\alpha,-\vec\beta,\lambda).
\end{eqnarray}

Due to the special dependence (\ref{phi-alpha-dep}) of $\psi_{\pm}$ and $\phi_{\pm}$ on $\vec\alpha$,  
also the Fourier transforms 
$\tilde K^{\pm}_{\pm}(\vec\alpha,\vec\omega,\lambda)$ of 
$K^{\pm}_{\pm}(\vec\alpha,\vec\beta,\lambda)$ with respect to $\vec\beta$, depend on $\vec\alpha$ only at the exponents:
\beq
\label{triangula2}
\ba{l}
\tilde K^{\pm}_{\pm}(\vec\alpha,\vec\omega,\lambda)=
\left(\tilde K^{\pm 1}_{\pm}(\vec\omega,\lambda) \right)^{\alpha_1}
\left(\tilde K^{\pm 2}_{\pm}(\vec\omega,\lambda) \right)^{\alpha_2},
\ea
\eeq
yielding the following Fourier representations of the kernels $K^{\pm}_{\pm}$:
\beq
\label{FT-K}
\ba{l}
K^{\pm}_{+}(\vec\alpha,\vec\beta,\lambda)=\int_{\RR^2}\frac{d\vec\omega}{(2\pi)^2}e^{i\vec\omega\cdot(\vec\alpha-\vec\beta)}
e^{\vec\alpha\cdot\vec\chi^{\mp}_{+}(\vec\omega,\lambda)}, \\
~~ \\
K^{\pm}_{-}(\vec\alpha,\vec\beta,\lambda)=\int_{\RR^2}\frac{d\vec\omega}{(2\pi)^2}e^{i\vec\omega\cdot(\vec\alpha-\vec\beta)}
e^{\vec\alpha\cdot\vec\chi^{\pm}_{-}(\vec\omega,\lambda)}, 
\ea
\eeq
where
\beq
\label{def_chi}
\vec\chi^{\mp}_{\pm}=(\chi^{\mp 1}_{\pm},\chi^{\mp 2}_{\pm}),~~\chi^{\mp j}_{\pm}=\ln \tilde K^{\pm j}_{\pm}-i\omega_j,~~j=1,2,
\eeq
and yielding also the representations of the analytic eigenfunctions $\vec\pi_{\pm}$ in terms of 
$\vec\varphi_{\pm}$ and $\vec\chi^{\mp}_{\pm}$:
\beq
\label{repr_pi}
\ba{l}
\vec\pi_{+}(\vec x,z,\lambda)=i\vec\varphi_{\pm}(\vec x,z,\lambda)+
\vec\chi^{\mp}_{+}(\vec\varphi_{\pm}(\vec x,z,\lambda),\lambda), \\
\vec\pi_{-}(\vec x,z,\lambda)=
i\vec\varphi_{\pm}(\vec x,z,\lambda)+\vec\chi^{\pm}_{-}(\vec\varphi_{\pm}(\vec x,z,\lambda),\lambda).
\ea
\eeq
The proof is the same as before; replacing the exponential representations (\ref{phi-exp}) and (\ref{phi-alpha-dep}b) 
of $\phi^{\pm}$ and $\psi_{\pm}$, f.i., in (\ref{K}a), one obtains the relation
\beq
e^{\vec\alpha\cdot\vec\pi_+}=\tilde K^{\pm}_{+}(\vec\alpha,\vec\varphi_{\pm},\lambda),
\eeq
proving that the dependence of $\tilde K^{\pm}_{+}$ on $\vec\alpha$ is only at the exponent. 
In addition, using the expression of $\tilde K^{\pm}_{\pm}$ in terms of  
$\vec\chi^{\mp}_{+}$, one obtains (\ref{repr_pi}a).  

We have clarified the impact of the ring property on the kernels $K^{\pm}_{\pm}$. The analyticity properties 
of the $z=\pm\infty$ asymptotics of $\psi_{\pm}$, discussed in the previous section, 
imply instead the following triangular structures:
\beq
\label{triangula1}
\ba{l}
K^{\pm}_{+}(\vec\alpha,\vec\beta,\lambda)=
\delta(\vec\alpha -\vec\beta)+\theta (\pm (\alpha_1-\beta_1))\check K^{\pm}_{+}(\vec\alpha,\vec\beta,\lambda), \\
K^{\pm}_{-}(\vec\alpha,\vec\beta,\lambda)=
\delta(\vec\alpha -\vec\beta)+\theta (\mp (\alpha_1-\beta_1))\check K^{\pm}_{-}(\vec\alpha,\vec\beta,\lambda)
\ea
\eeq
(the same features were observed in \cite{MZ}). Therefore the linear integral equation  (\ref{K-S}) describes  
a factorization problem allowing to construct, in principle, the kernels $K^{\pm}_{\pm}$ in terms of the 
given scattering datum $S$. 

Comparing (\ref{triangula1}) and (\ref{triangula2}), one infers that $\tilde K^{-}_{+}$, $\tilde K^{+}_{-}$ 
are analytic in the upper half $\omega^1$ - plane and $\tilde K^{+}_{+}$, $\tilde K^{-}_{-}$ 
are analytic in the lower half $\omega^1$ - plane. Since these analyticity properties are valid 
$\forall~\vec\alpha\in\RR^2$, it follows that  
$\tilde K^{\pm 1}_{\pm},~\tilde K^{\pm 2}_{\pm}$ in (\ref{triangula2}) cannot have poles and zeroes in their 
analyticity domains. Therefore also ${\chi^{+}_{\pm}}^{1,2}(\vec\omega,\lambda)$ and 
${\chi^{-}_{\pm}}^{1,2}(\vec\omega,\lambda)$ are analytic, rispectively, in the upper and lower half $\omega_1$ - plane, 
with asymptotics:
\beq
\ba{l}
{\chi^{\pm}_{\pm}}^{1,2}(\vec\omega,\lambda)\to 0,~~|\vec\omega |^2\to\infty,~\lambda\ne 0.
\ea
\eeq   
At last, plugging the Fourier representations (\ref{FT-S}) and (\ref{FT-K}) into 
the integral equation (\ref{K-S}), 
one obtains the following basic equations:
\beq
\label{RH}
\ba{l}
\vec\chi^{+}_+(\vec\omega,\lambda)-\vec\chi^{-}_+(\vec\omega+\vec \Delta (\vec\omega,\lambda),\lambda)=
i\vec\Delta^{1,2}(\vec\omega,\lambda),~~\\
\vec\chi^{-}_-(\vec\omega,\lambda)-\vec\chi^{+}_-(\vec\omega+\vec \Delta (\vec\omega,\lambda),\lambda)=
i\vec\Delta^{1,2}(\vec\omega,\lambda),~~
\vec\omega\in\RR^2,~~\vec\Delta\in\RR^2,~\lambda\in\RR,
\ea
\eeq 
wich must be viewed as ``inhomogeneous Riemann - Hilbert problems on the line Im $\omega^1=0$, with the shift 
$\Delta^1(\vec\omega,\lambda)$'' \cite{Mu,Ga} (the real variables $\omega^2$ and $\lambda$ appear just as parameters). Given 
$\vec\Delta(\vec\omega,\lambda)$, namely, given the shift and the inhomogenuity in (\ref{RH}), one can construct, 
in principle, the unique solutions $\vec\chi^{\pm}_{\pm}$ of these Riemann - Hilbert problems, 
expressed in terms of the following linear integral equations of Fredholm type : 
\beq
\ba{l}
{\vec\chi^{+}_+}(\omega^1,\omega^2,\lambda)+
\frac{1}{2\pi i}\int_{\RR}d\tau M(\omega^1,\tau,\lambda){\vec\chi^{+}_+}(\tau,\omega^2,\lambda)+
\vec L(\omega^1,\omega^2,\lambda)=
\vec 0, \\
{\vec\chi^{-}_-}(\omega^1,\omega^2,\lambda)-
\frac{1}{2\pi i}\int_{\RR}d\tau M(\omega^1,\tau,\lambda){\vec\chi^{-}_-}(\tau,\omega^2,\lambda)
+\vec L(\omega^1,\omega^2,\lambda)=
\vec 0,
\ea
\eeq 
where  
\beq
\ba{l}
M(\omega^1,\tau,\omega^2,\lambda)=\frac{\partial s^1_+(\tau,\omega^2,\lambda)/\partial\tau}
{s^1_+(\tau,\omega^2,\lambda)-s^1_+(\omega^1,\omega^2,\lambda)}-\frac{1}{\tau-\omega^1}, \\
\vec L(\omega^1,\omega^2,\lambda)=-i\vec\Delta(\omega^1,\omega^2,\lambda)+
\frac{1}{2\pi}P\int_{\RR}d\tau\frac{\partial s^1_+(\tau,\omega^2,\lambda)/\partial\tau}
{s^1_+(\tau,\omega^2,\lambda)-s^1_+(\omega^1,\omega^2,\lambda)}\vec\Delta(\tau,\omega^2,\lambda)  \\
s^1_+(\omega^1,\omega^2,\lambda)=\omega^1+\Delta^1(\omega^1,\omega^2,\lambda).
\ea
\eeq

\vskip 5pt
\noindent
{\bf The inverse problem}.    
Once $\vec\chi^{\pm}_{\pm}$ are known, one reconstructs the eigenfunction $\vec\varphi_-$ solving the 
following nonlinear integral equation
\beq
\label{phi-int-equ2}
i\vec\varphi_-(\vec x,z,\lambda)+\hat P^-_{\lambda}\vec\chi^+_+(\vec\varphi_-(\vec x,z,\lambda),\lambda)+
\hat P^+_{\lambda}\vec\chi^-_-(\vec\varphi_-(\vec x,z,\lambda),\lambda)=i\vec X, 
\eeq
where $\hat P^{\pm}_{\lambda}$ are respectively the $(+)$ and $(-)$ analyticity 
projectors with respect to $\lambda$:
\beq
\hat P^{\pm}_{\lambda}=\pm\frac{1}{2\pi i}\int_{\RR}\frac{d\lambda'}{\lambda'-\lambda \mp i\epsilon}\cdot
\eeq
Equation (\ref{phi-int-equ2}) is obtained applying $\hat P^-_{\lambda}$ and $\hat P^+_{\lambda}$ respectively 
to the analytic (and decreasing at $\lambda=\infty$) 
expressions $(\vec\pi_+-i\vec X)$ and $(\vec\pi_--i\vec X)$, subtracting the resulting equations, and   
using (\ref{repr_pi}). 

From the solution of (\ref{phi-int-equ2}), one reconstruct the potentials $\vec u_1$ using the formulas
\beq
\label{u_inv2}
\vec u_1(\vec x,z)=i\partial_x\displaystyle\lim_{\lambda\to\infty}
{\left(\lambda\big[i(\vec\varphi_-(\vec x,z,\lambda)-\vec X)+
\vec\chi^+_{+}(\vec\varphi_-(\vec x,z,\lambda),\lambda)\big]\right)},
\eeq
consequence of (\ref{u-Q}) and (\ref{asympt_pi}).

Therefore we have constructed an IST which involves a nonlinear step: the 
nonlinear equation (\ref{phi-int-equ2}) of the inverse problem. The above results can 
be summarized as follows. 

\vskip 5pt
\noindent
{\it Nonlinear IST scheme}. The direct scattering is the mapping from the potential $\vec u_1(\vec x,z)$ to the scattering 
vector $\vec\Delta(\vec\omega,\lambda)$. Knowing $\vec\Delta(\vec\omega,\lambda)$, one constructs the vectors 
$\vec\chi^+_+(\vec\omega,\lambda)$ 
and $\vec\chi^-_-(\vec\omega,\lambda)$ solving the linear RH problems with shift (\ref{RH}). In the inverse problem, 
from the knowledge of  
$\vec\chi^+_+(\vec\omega,\lambda),~\vec\chi^-_-(\vec\omega,\lambda)$, one solves 
the nonlinear integral equation (\ref{phi-int-equ2}) for $\vec\varphi_-(\vec x,z,\lambda)$ and, finally, one 
reconstructs the potentials $\vec u_1(\vec x,z)$ from (\ref{u_inv2}). 

It is quite remarkable that the inverse problem becomes linear, if expressed in terms of the 
Jost and analytic eigenfunctions $\phi^{\pm}$ and $\psi_{\pm}$. Indeed, from the solutions $\vec\chi^{\pm}_{\pm}$ 
of the RH problems (\ref{RH}) one can construct, via (\ref{FT-K}), the factorization 
kernels $K^{\pm}_{\pm}$. Known $K^{-}_{+},K^{-}_{-}$, one 
reconstructs the Jost eigenfunction $\phi^-$ by solving the following {\it linear} integral equation: 
\beq
\label{phi-int-equ}
\ba{l}
\phi^-(\vec\alpha,\lambda)e^{-i\vec\alpha\cdot\vec X}+
\hat P^{+}_{\lambda}
\int_{\RR^2}d\vec\beta \theta (-(\alpha_1-\beta_1))\check K^{-}_{-}(\vec\alpha,\vec\beta,\lambda)
\phi^-(\vec\beta,\lambda)e^{-i\vec\alpha\cdot\vec X} + \\
~~ \\
\hat P^{-}_{\lambda}
\int_{\RR^2}d\vec\beta \theta (-(\alpha_1-\beta_1))\check K^{-}_{+}(\vec\alpha,\vec\beta,\lambda)
\phi^-(\vec\beta,\lambda)e^{-i\vec\alpha\cdot\vec X}=1.
\ea
\eeq
This equation, in which we have omitted, for simplicity, the parametric dependence on $(\vec x,z)$, is consequence 
of the analyticity properties of $\psi_{\pm}$, and is obtained multiplying the equations
\beq
\label{dressing}
\ba{l}
\psi_{+}(\vec x,z;\vec \alpha,\lambda)=
\int_{\RR^2}d\vec\beta K^{-}_{+}(\vec\alpha,\vec\beta,\lambda)\phi^{-}(\vec x,z;\vec \beta,\lambda), \\
\psi_{-}(\vec x,z;\vec \alpha,\lambda)=
\int_{\RR^2}d\vec\beta K^{-}_{-}(\vec\alpha,\vec\beta,\lambda)\phi^{-}(\vec x,z;\vec \beta,\lambda)
\ea
\eeq
in (\ref{K}) by $e^{-i\vec\alpha\cdot\vec X}$, subtracting $1$, applying respectively $\hat P^{-}_{\lambda}$ 
and $\hat P^{+}_{\lambda}$,  
and adding the resulting equations. This inversion procedure has been already presented in \cite{Manakov1}. 

Once $\phi^{\pm}$ are known and, via (\ref{dressing}), $\psi^{\pm}$ are also known, the potential $\vec u_1$ 
is reconstructed, via (\ref{asympt1}) and (\ref{u-Q}), through the formulae:
\beq
\label{u_inv1}
\ba{l}
u^1_1(\vec x,z)=i\partial_x\left.\displaystyle\lim_{\lambda\to\infty}
{\lambda\left(\psi_{\pm}e^{-i\vec\alpha\cdot\vec X}-1\right)}\right|_{\vec\alpha=(1,0)}, \\
u^2_1(\vec x,z)=i\partial_x\left.\displaystyle\lim_{\lambda\to\infty}
{\lambda\left(\psi_{\pm}e^{-i\vec\alpha\cdot\vec X}-1\right)}\right|_{\vec\alpha=(0,1)}.
\ea
\eeq
This second version of the IST consists of linear steps only, and can be summarized as follows.  

\vskip 5pt
\noindent
{\it Linear IST scheme}. The direct scattering is the 
mapping from the potential $\vec u_1(\vec x,z)$ to the scattering 
vector $\vec\Delta(\vec\omega,\lambda)$. Knowing $\vec\Delta(\vec\omega,\lambda)$, one constructs the vectors 
$\vec\chi^+_+(\vec\omega,\lambda)$ and $\vec\chi^-_-(\vec\omega,\lambda)$ solving the linear RH problems with shift 
(\ref{RH}). Then, using the Fourier representations (\ref{FT-K}), one obtains the kernels $K^-_-$ and $K^-_+$. 
In the inverse problem, 
the Jost eigenfunction is then reconstructed through the linear integral equation (\ref{phi-int-equ}), the 
analytic eigenfunctions via (\ref{dressing}), and the potentials via (\ref{u_inv1}).

\vskip 20pt
\noindent
{\bf The small field limit}. In the small field limit $|u^{1,2}_1|=O(\epsilon)<<1$, the scattering vector 
$\vec\Delta$ is expressed in terms of $\vec u_1$ in the following way:
\beq
\vec\Delta(\vec\omega,\lambda)=\int_{\RR}dz\vec u_1(\omega^1+\lambda z,\omega^2,z)+O(\epsilon^2),
\eeq
while the Riemann - Hilbert problems with shift (\ref{RH}) reduce to the scalar Riemann - Hilbert problems:
\beq
\vec\chi^{+}_+(\vec\omega,\lambda)-\vec\chi^{-}_+(\vec\omega,\lambda)=
\vec\chi^{-}_-(\vec\omega,\lambda)-\vec\chi^{+}_-(\vec\omega,\lambda)=i\vec\Delta(\vec\omega,\lambda)+O(\epsilon^2),
\eeq
entailing
\beq
\label{linlim3}
\ba{l}
\vec\chi^{\pm}_+(\vec\omega,\lambda)=\frac{1}{2\pi}\int_{\RR^2}\frac{d\xi dz}{\xi-(\omega_1\pm i0)}
\vec u_1(\xi+\lambda z,\omega_2,z)+O(\epsilon^2), \\
\vec\chi^{\pm}_-(\vec\omega,\lambda)=-\frac{1}{2\pi}\int_{\RR^2}\frac{d\xi dz}{\xi-(\omega_1\pm i0)}
\vec u_1(\xi+\lambda z,\omega_2,z)+O(\epsilon^2).
\ea
\eeq
At last, the small field limit of the inverse problem equations  (\ref{phi-int-equ}) and (\ref{phi-int-equ2})  
reads 
\beq
i\big(\vec\varphi_-(\vec x,z,\lambda)-\vec X\big)+
\hat P^+_{\lambda}\vec\chi^{-}_-(\vec X,\lambda)+\hat P^-_{\lambda}\vec\chi^{+}_+(\vec X,\lambda)=O(\epsilon^2).
\eeq

\vskip 20pt
\noindent
{\bf The t-evolution}. The time evolution is introduced in the usual way. We observe that the Jost eigenfunction 
$\phi^+$ solves the 
equation  $\hat L_2\phi^+=i\lambda \alpha_2\phi^+$ which, evaluated at $z=-\infty$ using also (\ref{S}), leads to
\beq
\label{S(t)}
S(\vec\alpha,\vec\beta,\lambda,t)=S(\vec\alpha,\vec\beta,\lambda,0)e^{i\lambda(\alpha_2-\beta_2)t},
\eeq
From (\ref{S(t)}) and (\ref{K-S}) one infers the same elementary $t$-dependence for $K^{\pm}_{\pm}$: 
\beq
\label{K(t)}
K^{\pm}_{\pm}(\vec\alpha,\vec\beta,\lambda,t)=K^{\pm}_{\pm}(\vec\alpha,\vec\beta,\lambda,0)e^{i\lambda(\alpha_2-\beta_2)t}.
\eeq
At last, using (\ref{S(t)}) and the Fourier representation (\ref{FT-S}) of $S$, one obtains the 
$t$-evolution of the scattering vector $\vec\Delta$: 
\beq
\label{Delta(t)}
\vec\Delta(\vec\omega,\lambda,t)=\vec\Delta\big(\vec\omega-(0,\lambda)t,\lambda\big).
\eeq

\vskip 20pt
\noindent
{\bf The heavenly reduction}. The heavenly equation corresponds to the Hamiltonian reduction 
(\ref{red-u}). In this case, if $f_{1,2}$ are solutions of $\hat L_if_{1,2}=0$, 
then the Poisson bracket (\ref{PB}) of $f_{1,2}$ is also a solution: $\hat L_i(\{f_1,f_2\}_{\vec x})=0$. Therefore 
the solution space of the equation $\hat L_1f=0$ is not only a ring, but also a Lie algebra, with Lie bracket 
given by the Poisson bracket (\ref{PB}). This result 
follows immediately from equation (\ref{HamEqu}a) and from the Jacobi identity.

Under the hamiltonian reduction (\ref{red-u}), equation (\ref{dyn}) gets the form 
\begin{eqnarray}
\frac{d\vec r}{d z} =\{\vec r, H_1 + \lambda r_2\}_{\vec r}.
\end{eqnarray}
Thus the $z$-evolution of $\vec r$ is canonical, i.e. it preserves the Poisson brackets of $r_1(z)$ and $r_2(z)$. So, comparing 
eqs.(\ref{dyn}b) and  (\ref{asymptr+-}), one obtains the following constraint for the scattering 
vector $\vec\Delta$: 
\beq
\label{red-r}
\left\{ s^1_+,s^2_+\right\}_{\vec\omega}=
\left|\frac{\partial (s^1_+,s^2_+)}{\partial (\omega^1,\omega^2)}\right|=1.
\eeq 
\section{Conclusions and open problems}
In this paper we have constructed the IST for   
multidimensional vector fields, and we have used it to solve the Cauchy problem for the heavenly equation. 
Although these results have been derived in the particular case (\ref{N=2}), directly connected to the heavenly equation, 
they can be extended in a straightforward way to the general vector fields (\ref{L1L2}) and to the $(4+N)$ dimensional 
PDEs (\ref{quasilin-u}). 

Interesting open problems under present investigation are: i) the transition from the above formal results to rigorous ones. 
ii) The study of the spectral mechanisms 
(if any) which could cause a breaking of the initial profile for the PDEs under investigation (a typical feature of    
quasi-linear PDEs). iii) The costruction of explicit solutions within this spectral formalism.

\end{document}